\documentclass[aps,prl,showpacs,noshowkeys,amsmath,amssymb,amsfonts,superscriptaddress,reprint]{revtex4-1}
\usepackage{graphicx}
\usepackage{dcolumn}
\usepackage{bm}
\usepackage{color}
\usepackage{ulem} 
\begin{document}
\title{Collective Directional Locking of Colloidal Monolayers on a Periodic Substrate}
\author{Ralph L. Stoop}
\affiliation{Departament de F\'{i}sica de la Mat\`{e}ria Condensada, Universitat de Barcelona, Barcelona, Spain}
\author{Arthur V. Straube}
\affiliation{Department of Mathematics and Computer Science,
Freie Universit\"at Berlin, Berlin, Germany}
\affiliation{Group ``Dynamics of Complex Materials'',
Zuse Institute Berlin, Berlin, Germany}
\affiliation{Departament de F\'{i}sica de la Mat\`{e}ria Condensada, Universitat de Barcelona, Barcelona, Spain}
\author{Tom H Johansen}
\affiliation{Department of Physics, University of Oslo, P. O. Box 1048 Blindern, 0316 Oslo, Norway}
\affiliation{Institute for Superconducting and Electronic Materials, University of Wollongong, Northfields Avenue, Wollongong, NSW 2522, Australia}
\author{Pietro Tierno}
\email{ptierno@ub.edu}
\affiliation{Departament de F\'{i}sica de la Mat\`{e}ria Condensada, Universitat de Barcelona, Barcelona, Spain}
\affiliation{Universitat de Barcelona Institute of Complex Systems (UBICS), Universitat de Barcelona, Barcelona, Spain}
\affiliation{Institut de Nanoci\`encia i Nanotecnologia, IN$^2$UB, Universitat de Barcelona, Barcelona, Spain.}
\date{\today}
\begin{abstract}
We investigate the directional locking effects 
that arise when 
a monolayer of paramagnetic colloidal particles is driven across a triangular lattice of magnetic bubbles. We use an external rotating magnetic field to generate a two dimensional traveling wave ratchet forcing the transport of particles along 
a direction that intersects two crystallographic axes of the lattice. 
We find that, while single particles show no preferred direction, 
collective effects induce transversal current and directional locking at high density via a spontaneous symmetry breaking. 
The colloidal current may be polarized via an additional bias field that makes one transport direction energetically preferred. 
\end{abstract}
\pacs{05.60.Cd, 82.70.Dd, 75.70.Kw}
\maketitle
{\it Introduction.} Understanding the dynamic states that emerge
from interacting particles driven across periodic potentials
is relevant for a great variety of 
condensed matter systems~\cite{Han09,Van13,Cha17,Rei17}, ranging from electron scattering~\cite{Kue10} 
to vortices in superconductors~\cite{Bae95,Har96},
artificial spin ice~\cite{Yon18}, 
skyrmions~\cite{Rei18}, granular~\cite{Nit14} and active matter~\cite{Bec16}. 
Beside the fundamental insights, controlling the flow 
of microscale systems across ordered landscapes can have direct 
technological applications such us particle sorting and fractionation~\cite{Reic04,Lot04,Lac05,Spe09,Spe10,Eic10,Reguera2012}, 
processes that are relevant to both microfluidics~\cite{Mac03}
and biotechnology~\cite{Dav06}.

An astonishing effect that arises 
when the particles are driven through 
a regular array of obstacles or potential wells
is directional locking~\cite{Kor02,Kop10}. 
Such phenomenon occurs 
when these particles move along periodic trajectories 
that are commensurate with the underlying lattice,
in general not aligned with the direction of the driving force.
An experimentally accessible model system that 
allows investigating directional locking in real time and space, is based on the use of microscopic colloidal particles. This result from our ability to develop periodic potentials 
at the micro-scale using a variety of external means.
In previous experimental realizations 
directional locking has received much attention   
at the level of single particles,
and it manifested with the presence of 
a Devil's staircase structure
in the particle migration angle~\cite{Kor02,Bal09}. 
More recently, orientational and directional locking 
were observed 
for stiff clusters of microscopic particles driven by gravity across  
a periodic array of holes, and the resulting complex dynamics followed from competing symmetries between the two crystalline surfaces~\cite{Cao2019}.

For disperse and interacting multi-body systems, 
directional locking effect may emerge even in the most symmetric situation, i.e. when the particles are driven exactly at the 
middle of two crystallographic axes of the lattice.  
A related example in condensed matter is given by an ensemble of  
vortices in high T$_c$ superconductors forced to move across a honeycomb 
lattice of pinning sites. 
Numerical simulations of the vortex system predict the emergence of a transversal  current which result from the formation of dimeric state in the interstitial region of the lattice ~\cite{Rei08}.
We also note that directional locking effects have been recently reported on quasiperipodic substrates in experiments~\cite{Boh12} and on quasicrystalline substrates in simulations~\cite{Rei11}, although these studies have
been limited to fixed density of the interacting particles. 

Here we experimentally observe  
collective, density-dependent locking effects and transversal currents 
when a monolayer of paramagnetic colloidal particles is driven across a 
lattice of cylindrical ferromagnetic domains.
We show that, when 
the particles are driven 
along a direction that intersects two crystallographic axes, 
collective interactions
polarize the particle current along a preferred direction, 
not aligned along the driving one. 
Such direction results from a spontaneous symmetry breaking, and the corresponding velocity density curve displays a bifurcation diagram with two branches that can be destabilized via an additional bias field. Although directional
locking induced by collective interactions has been predicted 
in different theoretical works~\cite{Lee1999,Reichh04,Reic04,Rei08,Latimer2012},
we provide the first experimental 
observation of this general effect.

{\it Experiment.}
As magnetic substrate we use a ferrite garnet film (FGF) 
with uniaxial anisotropy which was
synthesized by liquid phase epitaxial growth~\cite{Tierno2009} on a ($111$) oriented
gadolinium
gallium garnet (GGG) substrate. The FGF
has composition Y$_{2.5}$Bi$_{0.5}$Fe$_{5-q}$Ga$_{q}$O$_{12}$ ($q=0.51$), 
\begin{figure}[t]
\begin{center}
\includegraphics[width=\columnwidth,keepaspectratio]{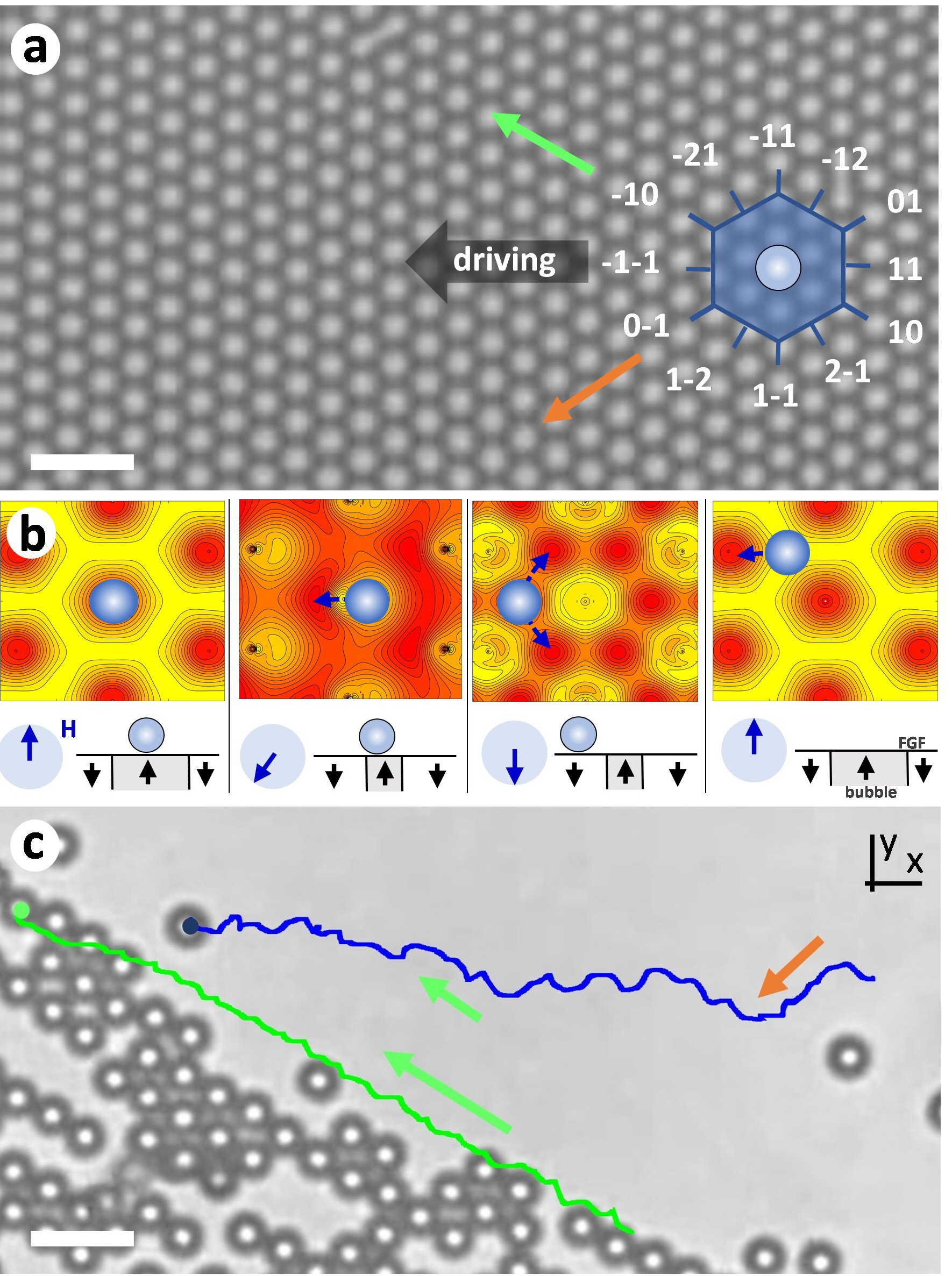}
\caption{
(a) Polarization microscope image showing a 
small portion of a magnetic bubble lattice with a schematic of the $12$ possible directions. The 
particles are driven toward left along the $-1-1$ direction.
; the basic lattice vectors are chosen as  $\bm{a}_{\pm}=(\sqrt{3}a/2,\pm a/2)$.
Scale bar is $10\rm{\mu m}$.
(b) Energy landscape of the magnetic lattice calculated at an elevation $z=0.7 a$ ($H_0=0.06 M_s$) for $t=0$, $t=3\pi/(4\omega)$, $t=\pi/\omega$ and $t=2\pi/\omega$ (from left to right, the particle moves from right to left). Schematics at the bottom shows one bubble in the FGF 
during the different phases of the applied field.
See~\cite{EPAPS} for details on the calculation.  
(c) Microscope image of the same region showing paramagnetic colloids driven 
towards left by a rotating field with 
$H_0=600 \rm{A m^{-1}}$, $\omega = 25.1\rm{rad s^{-1}}$
and $\beta=-1/3$. Superimposed are two trajectories of an individual particle (blue path) and one in a dense collection (green path). 
Scale bar is $10\rm{\mu m}$, see also Video1 in~\cite{EPAPS}.}
\label{figure1}
\end{center}
\end{figure}
saturation magnetization $M_s=1.3 \cdot 10^{4}\rm{ A m^{-1}}$
and it displays a triangular lattice of cylindrical domains  or "magnetic bubbles",  with diameter $D= 2.4 \rm{\mu m}$ at zero field, and lattice constant $a= 3.4 \rm{\mu m}$, see Fig.1(a). These cylindrical domains where generated using a strong gradient field $\nabla B \sim 0.2 \rm{T/m}$ perpendicular to the film before the experiments, and can be visualized by the polar Faraday effect, see Video1 in~\cite{EPAPS}. For low amplitudes, the size of the bubbles can be linearly tuned by a perpendicular field $\bm{H}^\text{ext}=H_z \hat{\bm{z}}$,
according to the relation $D=2 a\sqrt{(H_z/M_s+1)\frac{\sin{(\pi/3)}}{2\pi}}$.
On this film we deposit a water dispersion of paramagnetic colloidal particles 
with diameter $d=2.8 \rm{\mu m}$ (Dynabeads M-270) and magnetic volume susceptibility $\chi=0.4$.
In contrast to a previous work~\cite{Tie14},
we create here a periodic substrate composed of magnetic bubbles small enough to accommodate only one particle per unit cell. 
Before the experiments, the FGF is coated with a $1\rm{\mu m}$ thick layer of a  positive photoresist (AZ-1512, Microchem MA) using spin coating and UV photo-crosslinking \footnote{
This procedure avoids that the colloidal particles stick to the domain walls of the FGF due to their magnetic attraction. The gravitational length of the particles can be estimated as $h_g=k_B T/(V g \Delta \rho)= 61 nm$, thus much smaller than the particle size, and even in absence of attraction the system can be considered as effectively two-dimensional.}.
We use digital video microscopy~\cite{Cro96} to 
to keep track of the particle positions $(x_i(t),y_i(t))$ with $i=1...N$, and measure  
the instantaneous velocities $v_{x,y}$, here $v_x (t) = \frac{1}{N} \sum_i \frac{dx_i}{dt}$, and their mean values taken in the stationary regime \footnote{We use an upright microscope equipped with a CCD camera,
to record real-time videos of the system dynamics,
and analyze different subsets of a total field of view of $145 \times 108 \rm{\mu m}^2$.}.

{\it Particle transport.}
Over the FGF plate we displace the colloidal particles 
via an external  rotating magnetic field elliptically polarized in the 
$(\hat{x},\hat{z})$ plane,
$\bm{H}^\text{ext} = [-H_x \sin{(\omega t) \hat{\bm{e}}_x}+ H_z \cos{(\omega t) \hat{\bm{e}}_z}]$,
with frequency $\omega$ and amplitudes $(H_x,H_z)$. 
The total amplitude is given by $H_0=\sqrt{(H_x^2+H_z^2) /2}$
and the ellipticity parameter by 
$\beta = (H_x^2-H_z^2)/(H_x^2+H_z^2)$ with $\beta \in [-1, 1]$.
The applied field 
modulates the stray field $\bm{H}^\text{sub}$
on the FGF 
substrate, see~\cite{EPAPS} for details, 
and generates a two dimensional traveling wave ratchet
which drag along the particles in a deterministic way (thermal
noise is negligible). For $\omega < 31.4 \rm{rad s^{-1}}$, the transport is field synchronized, and the particles 
move ballistically with the speed of the traveling wave, $v_m=a \omega/(2 \pi)$, more details are given in~\cite{EPAPS}.
Throughout this work, we 
vary mainly the normalized particle density $\rho = (N d)/A$ within the observation area $A$, and keep $\omega = 25.1 \rm{rad s^{-1}}$ constant, which corresponds to a synchronous speed $\langle v_x \rangle = \sqrt{3} v_m /2 \simeq - 11 \rm{\mu m s^{-1}}$.
The field parameters are fixed to $H_0=600 \rm{A m^{-1}}$
and $\beta = -1/3$.

The single particle transport can be understood by calculating how the energy landscape of a paramagnetic colloid is
altered by the applied field during one field cycle, Fig.1(b).
When the field is parallel to the bubble magnetization,
$\bm{H}^\text{ext} = H_z \hat{\bm{z}}$ ($t=0$), the energy minimum, and thus the particle, is located at the center of the magnetic bubble.
For $t=\pi /(2 \omega)$ and $\bm{H}^\text{ext} = -H_x \hat{\bm{x}}$, the in-plane field $H_x$ deforms the landscape translating the 
particle to the interstitial region. When the field is anti-parallel 
to the bubble magnetization ($t=\pi /\omega$, $\bm{H}^\text{ext} = -H_z \hat{\bm{z}}$), it induces the nucleation of six equal energy minima at the interstitial region. 
For $t>\pi/\omega$, the interstitial minimum along the -1-1
direction start to disappear faster than the other two along the   
-10, and 0-1 crystallographic directions, as shown in FigS2. 
These two minima produce two equivalent pathways
along which the particle can be transported to reach 
one of the two closest magnetic bubbles ($t=\pi / \omega$). 
Since these pathways are energetically equivalent, the choice of the particle
is given by its initial position in the bubble minima, 
which is influenced by noise as thermal fluctuations.
As a result, the single particle transport 
displays no preference on the transversal ($\hat{\bm{y}}$) direction.

\begin{figure}[t]
\begin{center}
\includegraphics[width=\columnwidth,keepaspectratio]{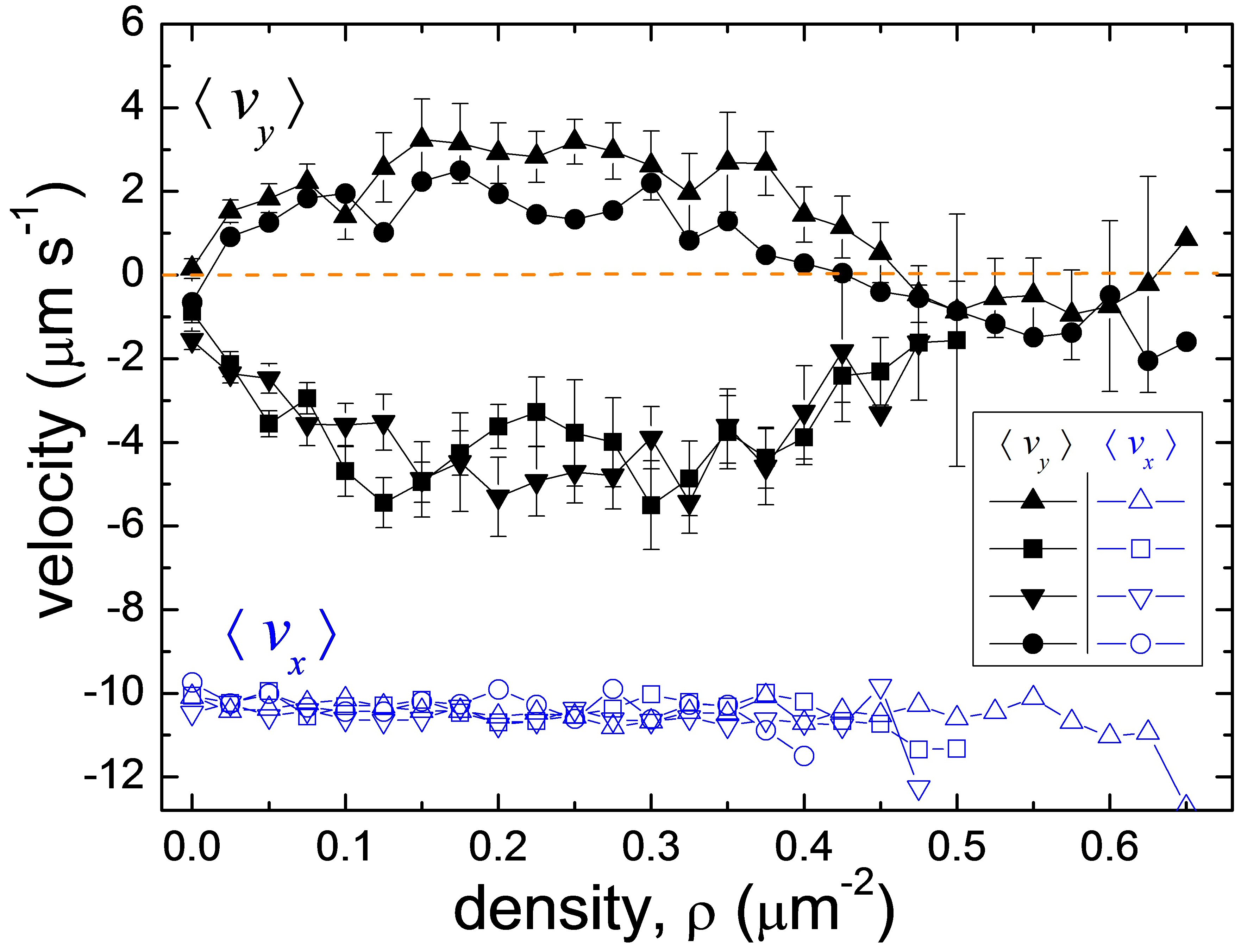}
\caption{Longitudinal velocity $\langle v_{x} \rangle$ (blue open symbols) and transversal one $\langle v_{y} \rangle$ (black filled symbols) versus normalized particle density $\rho$ for paramagnetic colloids driven toward left. Error bars are obtained from the
statistical analysis of different experimental realizations.}
\label{figure2}
\end{center}
\end{figure}

{\it Results.} To investigate directional locking effects we orient the magnetic substrate such that the driving direction ($\hat{x}<0$) is exactly between the two symmetry axes, $-10$ and $0-1$, Fig.1(a) \footnote{We note that if driven along a crystallographic axis, the particle transport is rather stable, and the dynamics shows a transition from a synchronous to asynchronous transport when $\omega$ is increased, as reported for a similar one-dimensional system~\cite{Str13}, see also Fig.S2 in~\cite{EPAPS}.}. 
In this situation, 
single particles display erratic trajectories, 
composed by an alternation of up and down interstitial jumps 
and a vanishing transversal velocity $\langle v_y \rangle=0$, Fig.1(c).
However, we find that for denser systems the colloidal current polarizes following one of the two crystallographic directions.  
For example, in Fig. 1(c) the dense collection of particles flows along the $-10$ direction ($\langle v_y \rangle >0$).
The observed phenomenon is rather robust, and lasts 
as long as the particle density $\rho$ 
or the applied field are kept constant. The chosen direction results from dynamically symmetry breaking: switching off and later on the rotating field $\bm{H}$
can induce transport along the complementary crystallographic direction, $0-1$, with a negative velocity 
$\langle v_y \rangle <0$. 
Moreover we find that along the chosen path, the colloidal flow is highly ordered, with few particle rearrangements. Now the particles form elongated and compact clusters which keep their shape during motion, as shown in Fig.S3 in~\cite{EPAPS},  
where we report the corresponding evolution of mean cluster size and the degree of clustering. 

\begin{figure}[b]
\begin{center}
\includegraphics[width=0.8\columnwidth,keepaspectratio]{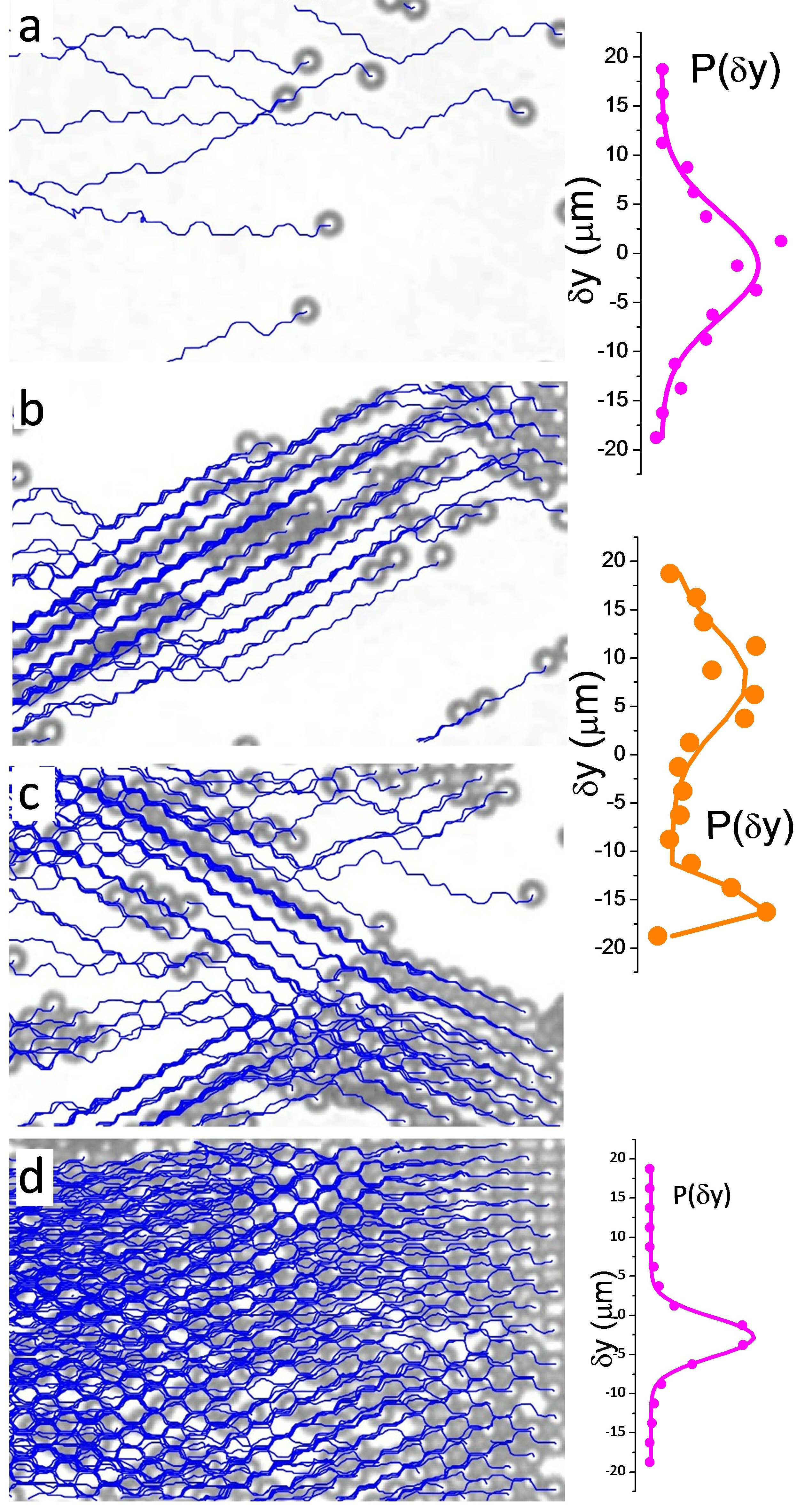}
\caption{Experimental images with superimposed particle trajectories for: (a) $\rho = 0.02$; (b) $\rho = 0.23$
(c) $\rho = 0.16$;
(d) $\rho = 0.58$.
Lateral graphs show the
distribution of the transversal displacement of the particles
$P(\delta y)$ for three different regions of the diagram 
with  $\rho < 0.025$ (top left),
$\rho \in [0.55,0.65]$ (bottom left) 
and $\rho \in [0.17,0.27]$ (right).
The $P(\delta y)$ at the top ($R^2=0.85$) and 
at the bottom ($R^2=0.99$) are fitted with a Gaussian distribution, while the middle curve is a bimodal fit ($R^2=0.82$), being  $R^2$ the coefficient of determination.}
\label{figure3}
\end{center}
\end{figure}

\begin{figure}[b]
\begin{center}
\includegraphics[width=\columnwidth,keepaspectratio]{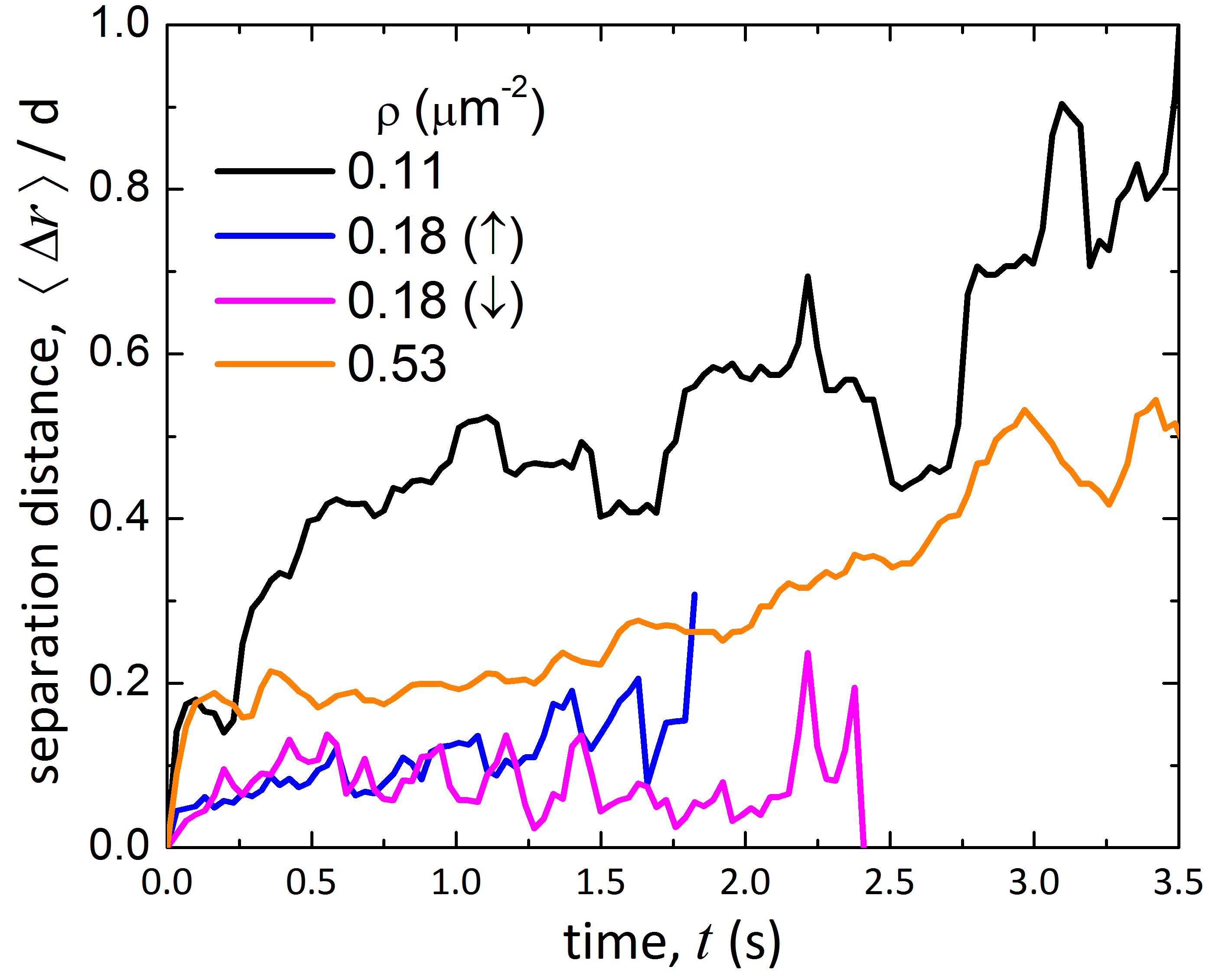}
\caption{Normalized average distance between nearest neighbours $\langle \Delta r \rangle$ versus time calculated for four densities.}
\label{figure4}
\end{center}
\end{figure}

Fig.2 shows the current density diagram measured in four different experiments, all of which conducted by transporting the paramagnetic colloids 
towards left, $\langle v_x \rangle<0$, corresponding images are illustrated in the four left panel of Fig.3. We measure a bifurcation diagram showing two nearly symmetric branches starting from $\rho \sim 0.05$,
and that reach a maximum separation speed 
of $\sim 8 \rm{\mu ms^{-1}}$ for $\rho \in [0.17,0.27]$. In all cases, 
the mean longitudinal velocity remains constant
to $\langle v_x \rangle=-10.3 \rm{\mu m s^{-1}}$. Above $\rho \sim 0.5$
the two branches merge and the bistability is lost for large clusters. 
In this situation, the colloidal particles form a percolating  network that covers the whole potential landscape. Even if the particle density is below close packing, the colloids have now nearest neighbors which impede the formation of polarized clusters moving along a defined crystallographic axis.
Further, we note that in the 
experimental data the two branches are not exactly centered 
at $\langle v_y \rangle=0$  indicating that the magnetic field distribution is not exactly symmetric, due to the presence of imperfections in the bubble lattice, as double magnetic domains. These imperfections create dislocation in the bubble array which may induce change in the crystallographic direction of the driven clusters, see also Fig.S3 in~\cite{EPAPS} for the case of Figs.3(b,c). We further measure
the distribution of lateral displacement $P(\delta y)$ for all set of data, as shown in Fig.3. Here we find that $P(\delta y)$
are Gaussian for both low ($\rho < 0.025$, Fig.3(a)) and high ($\rho \in [0.55,0.65]$, Fig.3(d)) densities where the branches merge. However at high density the width of the Gaussian ($\sigma = 5 \mu m$) reduces by half the value at low density ($\sigma = 10.6 \mu m$) due to the high concentration of particles, and the corresponding reduction of their lateral mobility. 
In contrast, for intermediate density ($\rho \in [0.17,0.27]$, Figs.3(b,c)) we find a bimodal distribution, which reflects the bistability reported in Fig.2. 

The transverse polarization produces a dynamical synchronization process between nearest propelling particles in the clusters. We quantify this effect in Fig.4 by measuring the average distance between nearest neighbor ($nn$) particles, $\langle \Delta r \rangle= \langle r_{ij}(t) - r_{ij}(0) \rangle_{nn}$, being $r_{ij}=|\bm{r}_i-\bm{r}_j|$ and $\bm{r}_i$ the position of particle $i$. For dilute density, the particles are far away and dipolar interactions are unable to form the clusters, thus $\langle \Delta r \rangle$ is observed to increase over with time in our experimental area. For intermediate densities ($\rho= 0.18$) the particles form elongated clusters, and the lateral jumps of the composing particles across the interstitial region (Fig.~\ref{figure1}(b)) are coherently synchronized. In this situation the average distance between two particles  is of the order of the lattice constant $a$, where $r_{ij} \sim a \, \forall t \, , \, \langle \Delta r \rangle \rightarrow 0$. At high density the directional locking is lost and the particles within the percolating structure display uncorrelated upward and downward lateral jumps. Thus, $\langle \Delta r \rangle$ raises but slowly due to the high particle density.

\begin{figure}[t]
\begin{center}
\includegraphics[width=\columnwidth,keepaspectratio]{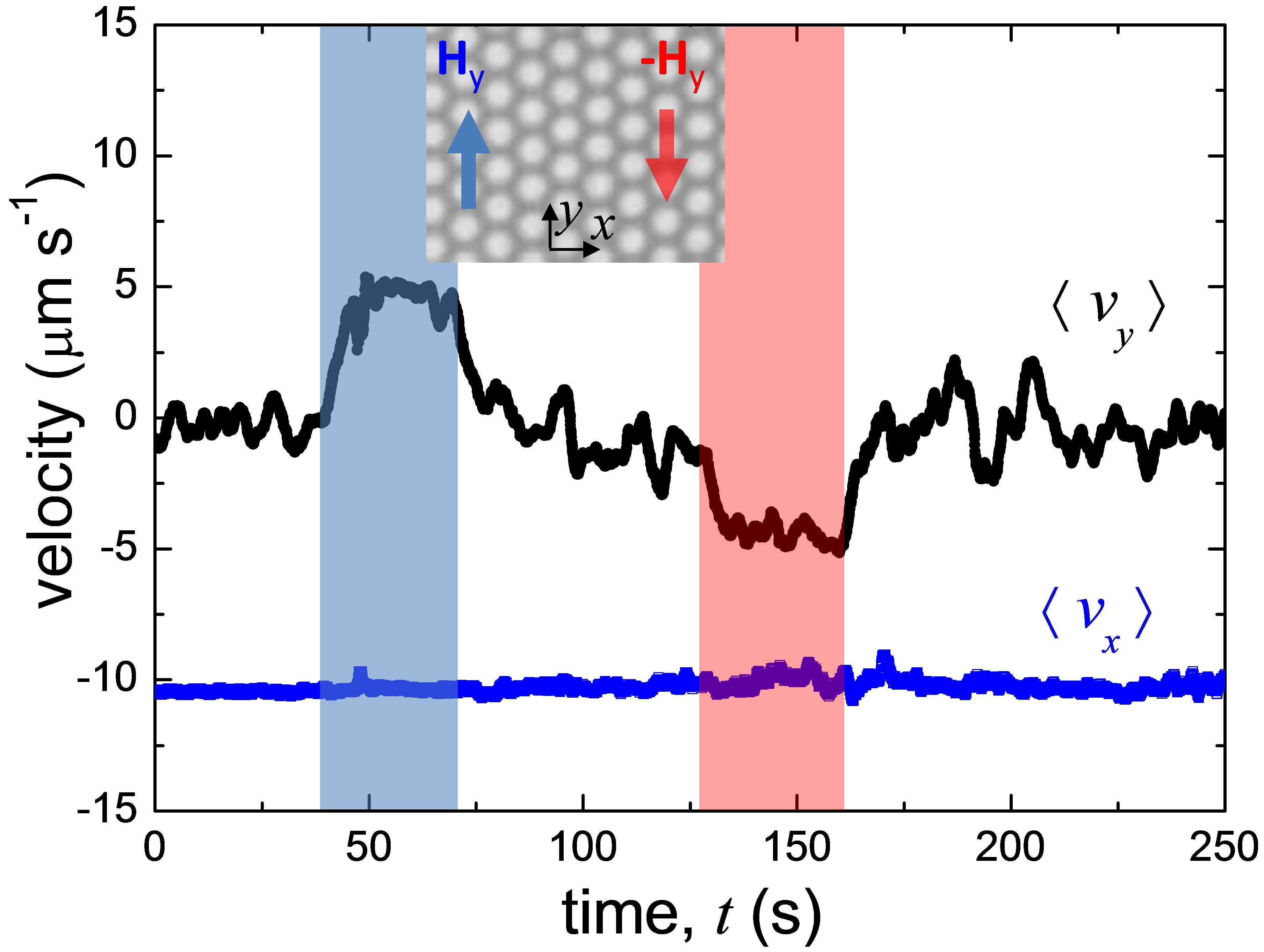}
\caption{Velocities 
versus time for a dilute system driven to the left, where density is slowly decreasing in time from $\rho = 0.06$ ($t=0$s) to $\rho = 0.06$ ($t=250$s). The shaded regions indicate the presence of an additional magnetic field $H_y=200 \rm{A m^{-1}}$ (blue) and $-H_y$ (pink). Right graph denotes the corresponding evolution of the normalized particle density $\rho$. 
Inset shows the direction of the applied bias fields with respect to the magnetic lattice.}
\label{figure5}
\end{center}
\end{figure}

The collective directional locking at intermediate $\rho$ results from the formation of compact clusters kept by attractive dipolar interactions. Via approximate expressions of the magnetic field, see~\cite{EPAPS} for details, the dipolar interaction between two particles with coordinates $\bm{r}_1=(x_1,y_1)$ and $\bm{r}_2=(x_2,y_2)=\bm{r}_1-\bm{a}_{\pm}$, i.e. separated by one lattice constant $a$ and aligned along a crystallographic axis, $\bm{a}_{\pm}=(\sqrt{3}a/2,\pm a/2)$ is reduced to
\begin{align}
U_\text{d}({\mathbf r}_{1}) \propto F_{xx} (1-3 \,\hat{x}_{12}^2)+F_{zz} -3 F_{xy} \,\hat{x}_{12}\, \hat{y}_{12}, 
\label{U-dip-approx}
\end{align}
where $F_{xy}= 2 H_x^\text{ext}(t) H_y^\text{sub}({\mathbf r}_1)$ and similarly for $F_{xx}$, $F_{zz}$, and $\hat{x}_{12}=\sqrt{3}/2$, $\hat{y}_{12}=\pm 1/2$. Potential (\ref{U-dip-approx}) has a hexagonal structure with attraction along the crystallographic axes. These interactions are caused by the interplay between stray field and external modulation while evolving in time, and is a complex and richer effect than that reported before~\cite{Str14}. Our model neglects hydrodynamic interactions suggesting that magnetic forces dominate the dynamics of the system.

For intermediate densities, the dynamical state of our driven monolayer is bistable, and switching between them can be controlled by an additional bias field $H_y$ superimposed to the driving one. We demonstrate this feature in Fig.5, where the particle velocities are plotted as a function of time for a low density case ($\rho \sim 0.1$) where $\langle v_y \rangle=0$, while $\langle v_x \rangle =-10 \rm{\mu m s^{-1}}$. The addition of a static, transversal component $-H_y$ ($H_y$) to the rotating field during two time intervals $t\in[36.5, 70.6]$s ($t\in [128.9,161.9]$s) polarizes the colloidal current along the $0-1$ ($-10$) direction with $\langle v_y \rangle =-4.9 \rm{\mu m s^{-1}}$ ($\langle v_y \rangle =4.6 \rm{\mu m s^{-1}}$)
while the longitudinal current remains practically unaffected. Inspection of how the bias changes the two energy landscapes, reveals that the applied field produces a preferred energetic path during the particle excursion process, creating interstitial wells with unequal depths and thus transversal rectification along the bias direction. The polarization effect enables to selectively control the transversal rectification process, 
in contrast to the spontaneously chosen transport 
path in absence of bias.
We also stress the fact that our colloidal current is generated via time-modulated homogeneous magnetic fields, and not via external field gradient as in magnetophoresis~\cite{Zbo15}, that would impede the observation and control of transversal current across the lattice. 

To conclude, we demonstrate spontaneous 
directional locking that emerges by collective interactions when 
a monolayer of colloidal particles is driven across a periodic substrate. Our results have been obtained with 
a specially prepared single crystal uniaxial 
ferromagnetic thin film, however they are more general 
and can be observed in other condensed matter systems 
characterized by interacting particles driven across corrugated 
landscape. Example of transverse response and bistability effects have been reported in the past for vortex matter in high T$_c$ superconductors~\cite{Rei08} or electron scattering
\cite{Wiersig2001}. 
Also, the understanding and control of 
sliding process between particles on corrugated 
substrate may be of interest 
for studying friction and adhesion phenomena~\cite{Vanossi2013}. Here we add a mesoscopic model system for locking effects where particle currents and fluctuations can be investigated in real time/space and controlled by an external field. 

\begin{acknowledgments}
We thank Thomas M. Fischer for inspiring discussions. 
R. L. S. acknowledges support from the Swiss National Science Foundation grant No. 172065. 
A. V. S. was partially supported by the Deutsche Forschungsgemeinschaft (DFG, German Research Foundation) through the grant SFB 1114, project C01 and from MATH${+}$: the Berlin Mathematics Research Center (under Germany's Excellence Strategy, EXC-2046/1 project ID: 390685689), project EF4-4. 
P. T. acknowledge support from the
ERC Grant "ENFORCE" (No. 811234), 
MINECO (FIS2016-78507-C2-2-P, ERC2018-092827) and  Generalitat de Catalunya under program "Icrea Academia".
\end{acknowledgments}
\bibliography{biblio}
\end{document}